\documentstyle[12pt,epsf]{article}
\input epsf
\begin{document}
\noindent
{\Large \bf Global and Local Measures of the Intrinsic Josephson Coupling in Tl$_2$Ba$_2$CuO$_6$ }  \\
\\
Nature {\bf 395} (1998) 360-362, {\em 24 september 1998 }
\\
\\
A.A. Tsvetkov*\$, D. van der Marel*, K.A. Moler\&, J. R. Kirtley+, J. L. de Boer*, A. Meetsma*, Z.F. Ren\#, 
N. Koleshnikov$\parallel$, D. Dulic*, A. Damascelli*, M. Gr\"uninger*, J. Sch\"utzmann*, J.W. van der Eb*, 
H. S. Somal*, and J.H. Wang\#. 
\\
{\em
* Materials Science Centre, University of Groningen, 9747 AG Groningen \\
\$ P. N. Lebedev Physical Institute, 117924 Moscow, Russia   \\
\& Department of Physics, Princeton University, Princeton, NJ 085544     \\
+ IBM T. J. Watson Research Center, P.O. Box 218, Yorktown Heights, NY 10598   \\
\# Department of Chemistry, Suny at Buffalo, Buffalo, NY 14260-3000    \\
$\parallel$ Institute of Solid State Physics, Russian Academy of Sciences, Chernogolovka  }
\\ 
One leading candidate theory of the high-temperature superconductors in the copper oxide systems  is the Inter-Layer 
Tunneling (ILT) mechanism[1]. In this model superconductivity is created by tunneling of electron pairs between the copper 
oxdide planes- contrasting with other models in which superconductivty first arises by electron pairing within each plane. The 
ILT model predicts that the superconducting condensation energy is approximately equal to the gain in kinetic energy of the 
electron pairs due to tunneling.  Both these energies can be determined independently[2-4], providing a quantitative test of the 
model.  The gain in kinetic energy of the electron pairs is related to the interlayer plasma energy, $\omega_J$, of electron pair 
oscillations, which can be measured using infrared spectroscopy. Direct imaging of magnetic flux vortices also provides a test[5], 
which is performed here on the same samples. In the high-temperature superconductor Tl$_2$Ba$_2$CuO$_6$, both the sample 
averaging optical probe and the local vortex imaging give a consistent value of $\omega_J$= 28 cm$^{-1}$ which, when combined with the 
condensation energy produces a discrepancy of at least an order of magnitude with deductions based on the ILT model.  \\
\begin{figure}
        \parbox[l]{160mm}{
        \leavevmode  
        \hbox{%
        \epsfxsize=100mm  
        \epsffile{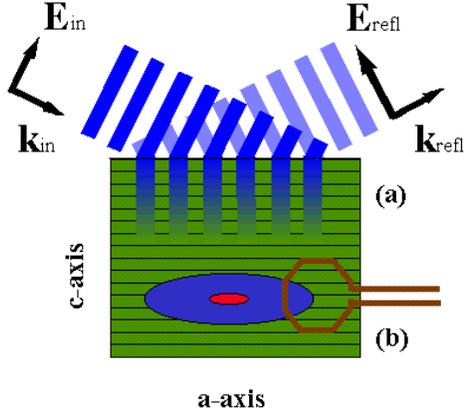} } }
        \caption{\em Schematic of the experimental techniques:  
        a) Grazing incidence reflectivity. The p-polarized light incident at 
        a grazing angle sets up a periodic electric field pattern, 
        which is polarized perpendicular to the sample surface, and 
        decays exponentially inside the solid.  
        b) Scanning SQUID microscopy. 
        The octagonal pickup loop detects the 
        magnetic flux perpendicular to the a-c face. }
\label{fig1}
\end{figure}
In the ILT model the normal state is different in nature from the traditional Landau Fermi liquid. As a result coherent transport of single charge 
carriers between the planes is strongly inhibited in the normal state. In the superconducting phase tunneling of pairs is possible, and the 
superconducting condensation energy (E$_{cond}$) in the ILT model is precisely the gain in kinetic energy (E$_J$) due to the tunneling of those pairs: E$_J$ 
= $\eta$E$_{cond}$. The number $\eta$ is of order 1 when ILT is the only active pairing mechanism[3]. 
With conventional mechanisms, although usually $\eta \ll 1$, 
there is no prediction for $\eta$ that is free from materials parameters. A crucial point in this discussion is, that both E$_{cond}$ and E$_J$ are experimentally 
accessible quantities, thus allowing the experimental verification of the ILT hypothesis. E$_{cond}$ can be measured 
from the specific heat[6]. E$_J$ can 
be determined by measuring the interlayer (Josephson) plasma frequency[7]. 
\\ For this work, we used two kinds of samples: single crystals and epitaxial thin films of Tl$_2$Ba$_2$CuO$_6$. The crystals have a transition temperature 
of 82K and transition width (10$\%-90\%$) of 13 K, as determined by bulk SQUID susceptibility. Using 4-circle X-ray diffraction we verified that the 
material belongs to the tetragonal {\em I4/mmm} space group, with (for the crystals) lattice parameters a=b=3.867$\AA$ , 
and c=23.223$\AA$ . The films 
have T$_c$=80K as determined by DC resistivity, and c=23.14 $\AA$ . Both types of samples have relatively large physical dimensions 
perpendicular to 
the c-axis, corresponding to the conducting copper oxide planes (50 mm$^2$ for the thin films, and 1 mm$^2$ for the crystals). They have small 
dimensions along the c-axis (~1 $\mu$m for the thin films, and ~50 $\mu$m for the crystals). 
\\ To determine the plasma resonance we measure the reflection coefficient of infrared 
radiation incident on the ab-plane at a large angle (80$^o$) 
with the surface normal[8]. A sketch of the experiment is presented in Fig. 1. In the case of the single crystals the reflected light drops below our 
detection limit if the wavelength exceeds 0.2 mm (i.e. for $\omega/2\pi c < 50 cm^{-1}$) due to diffraction. Using thin films we were able to extend this range 
to 20 cm$^{-1}$. The electric field vector of the radiation is chosen parallel to the plane of reflection, resulting in a large component 
perpendicular to 
the CuO$_2$ planes. This geometry allows absorption of the light by lattice vibrations and plasma-oscillations polarised 
perpendicular to the planes. 
%
\begin{figure}
        \parbox[l]{160mm}{
        \leavevmode  
        \hbox{%
        \epsfxsize=100mm  
        \epsffile{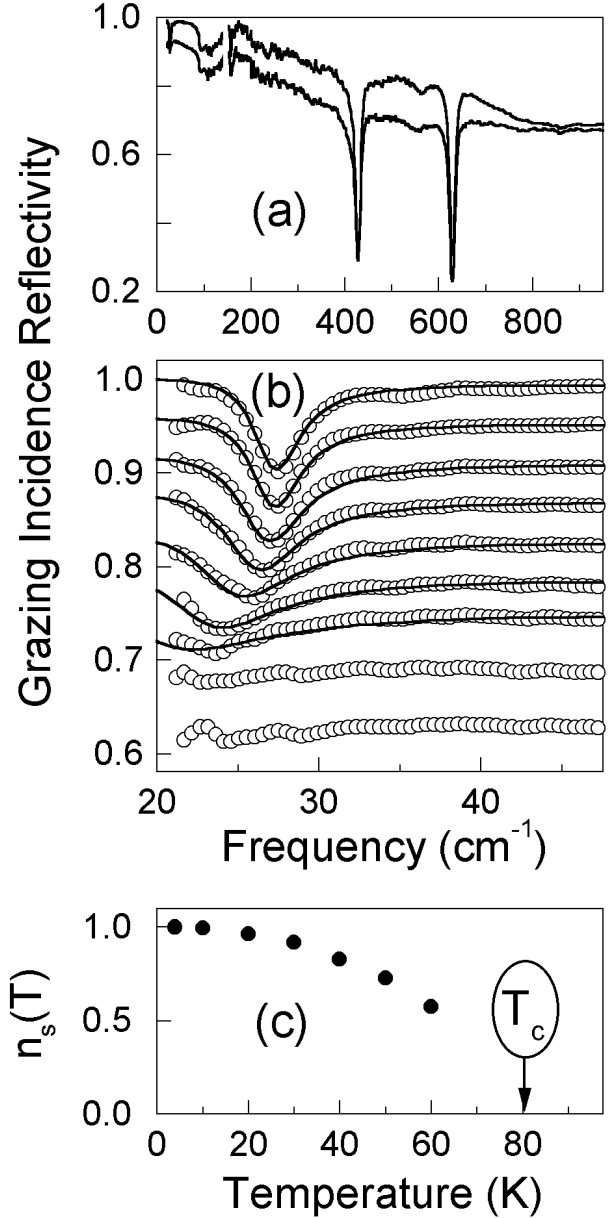} } }
        \caption{\em a) P-polarized reflectivity at 80$^o$ angle of incidence of Tl$_2$Ba$_2$CuO$_6$ 
        at 4 K (uppercurve) and 100 K (lower curve). Frequencies above (below) 150 cm$^{-1}$ 
        correspond to single crystal (thin film) data. 
        b) Thin film spectra on an 
        expanded frequency scale. From top to bottom: 4K, 10 K, 20 K, 30 K, 40 K, 50 K, 60 K, 75 K, 
        nd 90 K. The curves have been given incremental 3 percent vertical offsets for clarity. 
        The solid curves correspond to calculations as described in the text. 
        c) Temperature dependence of $n_s(T)=\omega_J(T)^2/\omega_J(4K)^2$, demonstrating that the 
        resonance frequency converges to zero at T$_c$. }
\label{fig2}
\end{figure}
\\In Fig. 2a we present the single crystal and thin film reflectivity for $\omega/2\pi$c respectively above and below 150 cm$^{-1}$. All prominent absorption lines 
for frequencies above 50 cm$^{-1}$ correspond to infrared active lattice vibrations, which show no strong temperature dependence. In the 4K 
spectrum we observe a clear absorption at 27.8 cm$^{-1}$. This resonance exhibits a strong red shift upon raising the temperature, as displayed in 
Fig. 2b. Above 70 K it has shifted outside our spectral window. In Fig. 2c we also present the temperature dependence 
of $\omega(T)^2/\omega(4K)^2$ of the 
resonance position. This temperature dependence extrapolates to zero at T$_c$, which indicates, that it is a plasma resonance of the paired charge 
carriers. We therefore attribute this absorption to a Josephson plasmon, a collective oscillation of the paired charge carriers 
perpendicular to the 
coupled superconducting planes[7]. For a purely electronic system the supercurrent density along the c-axis determines the Josephson resonance 
frequency, $c/\lambda_c$. Because in the present case the Josephson plasma resonance is located at a frequency below the infrared active lattice 
vibrations, the corresponding dynamical electric field is screened by the ions and the lattice vibrations, characterised by a dielectric constant ecs. 
As a result we observe the Josephson resonance at a reduced frequency $\omega_J = \epsilon_{cs}^{-1/2} c /\lambda_c$. We performed a full optical analysis of these 
spectra in the spectral range from 20 to 6000 cm$^{-1}$ using Fresnel's equations for oblique angle of incidence reflection of anisotropic optical 
media. This way we were able to extract the dielectric function ecs from our data. For frequencies below 40 cm$^{-1}$, $
\epsilon_{cs} = 11.3 \pm 0.5$. We 
therefore obtain $\lambda_c(4K)= 17.0 \pm 0.3 \mu m$. 
\begin{figure}
        \parbox[l]{160mm}{
        \leavevmode  
        \hbox{%
        \epsfxsize=100mm 
        \epsffile{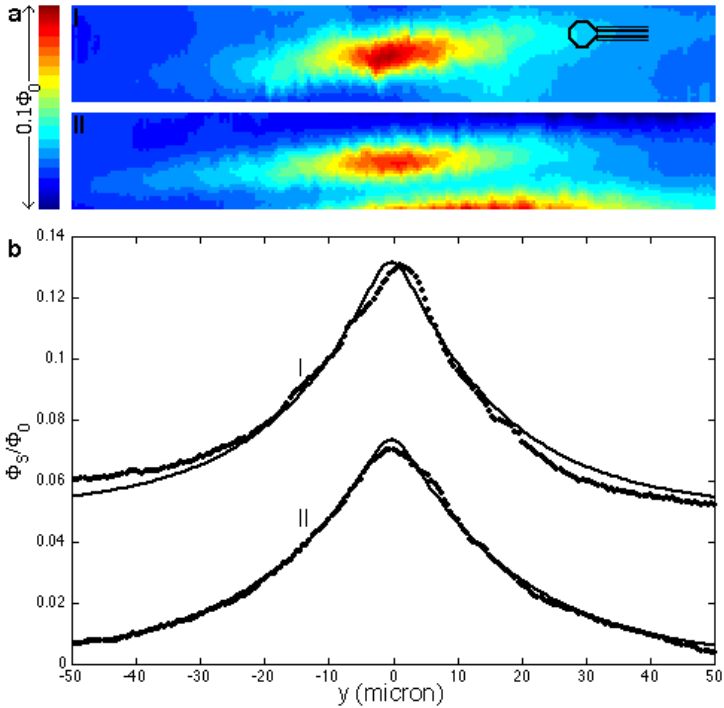} } }
        \caption{a) Images of the magnetic field perpendicular to an a-c face of a 
        Tl$_2$Ba$_2$CuO$_6$ single crystal in two different 
        locations, showing two different interlayer Josephson vortices. The dashed lines 
        indicate the longitudinal 
        cross-sections. Inset: sketch of the 4 mm octagonal pickup loop. 
        b) The flux through the SQUID pickup loop along 
        the longitudinal cross-sections. 
        The solid curves are fits which determine the c-axis penetration depths of these two 
        vortices to be (I) $\lambda_c = 17\pm 4$ mm and (II) $\lambda_c = 1\pm 1$ mm. 
        The straight lines indicate the extent of $\lambda_c$ for each vortex. }
\label{fig3}
\end{figure}
\\An independent experimental measure of the interlayer coupling is provided by a direct measurement of $\lambda_c$. Here we employ the fact that 
vortices which are oriented parallel to the planes, called interlayer Josephson vortices, have a characteristic size 
$\lambda_c$ along the planes and $\lambda_a$
perpendicular to the planes[9]. In order to determine $\lambda_c$ directly, we used a scanning Superconducting QUantum Interference Device (SQUID) 
microscope10 to map the magnetic fields perpendicular to an a-c face at 4 K (Fig. 1b) [5]. The crystal was cooled in a magnetically shielded 
cryostat with a residual magnetic field of a few milligauss, resulting in the presence of a few isolated trapped vortices (Fig. 3a). The jitter 
apparent in this image is due to the mechanical scanning mechanism used in our SQUID microscope. With an L = 4 mm octagonal pickup loop, 
the vortices were resolution-limited along the c direction ($\lambda_a \ll L$), but not along the a direction ($\lambda_c \ll L$). 
\\Fitting the longitudinal cross-sections (Fig. 3b) to the functional form for the magnetic fields of an interlayer Josephson vortex[9] convoluted with 
the shape of the pickup loop[5,10], gave the results $\lambda_c$ = 17 $\pm$ 4 mm and $\lambda_c$ = 19 $\pm$ 1 mm for these two vortices. The statistical error bars were 
determined using a criterion of doubling of the variance from the least-squares value, but systematic errors from the background and the shape 
of the pickup loop, and the effect of the surface on the shape of the vortex, which may be as large as 30
vortices in three pieces cut from a large single crystal, which was part of the mosaic used to make the measurements in Fig. 1a. The vortices in 
all three pieces confirm the plasma resonance frequency of ~ 28cm$^{-1}$. 
\\We are now ready to determine E$_J$ using the expression[7] $\epsilon_{cs}\hbar^2\omega_J^2=4\pi d a^{-2}e^{*2}E_J$, 
where $d$ is the distance between planes (11.6$\AA$ ), $a$ is the cell 
parameter (3.87$\AA$ ), $\hbar$ is the Planck constant, 
and $e^*=2e$ is the charge of the pairs. The result is $E_J=0.24 \mu eV$ per formula unit. For Tl$_2$Ba$_2$CuO$_6$ 
the measured value of E$_{cond}$ is 100 $\pm 20 \mu eV$ per formula unit[6]. Hence $\eta=E_J/E_{cond}= 0.0024 \pm 0.0005$, which is clearly at variance with the notion 
that condensation in the high-T$_c$ superconductors is due to the gain in kinetic energy of the pairs (E$_J$) in the superconducting state. 
\\Our observation of the same value of $\omega_J$ with a local probe (scanning SQUID) and a sample averaging probe (grazing reflectivity) demonstrates 
that we measure the intrinsic Josephson coupling, rather than a coupling determined by isolated metallurgical defects. The body of data 
presented in this paper provides strong support for the interpretation of both the Josephson plasma resonance and the interlayer Josephson 
vortices as intrinsic properties of Tl$_2$Ba$_2$CuO$_6$. One of the key predictions of the ILT model, 
that $\eta$=E$_J$/E$_{cond}$ is of order 1, is far outside the range of our experiments, which give $\eta$=0.0024. 
\\
{\em Received 13 March, accepted 29 june 1998 }
\\

{\bf Acknowledgements.} We thank M. Bhushan and M. Ketchen for assistance with the development of the scanning SQUID microscope. We gratefully acknowledge P. W. Anderson for 
stimulating comments, and P. Stamp for carefully reading the manuscript. KAM acknowledges the support of an R.H. Dicke postdoctoral fellowship. This work was supported by the 
Netherlands Foundation for Fundamental Research (FOM) with financial from the Nederlandse Organisatie voor Wetenschappelijk Onderzoek (NWO). AAT acknowledges the support 
by NWO, RVBR and the Russian Superconductivity Program. The work performed at SUNY-Buffalo is partly supported by NYSERDA. 
\\
\\
Correspondence and requests for materials should be addressed to Prof. Dr. D. van der Marel, electronic mail: D.van.der.marel@ phys.rug.nl 

\end{document}